\documentclass{PoS}
\newcommand{\gtsim}{\mbox{{\raisebox{-0.4ex}{$\stackrel{>}{{\scriptstyle\sim}}
$}}}}
\newcommand{\ltsim}{\mbox{{\raisebox{-0.4ex}{$\stackrel{<}{{\scriptstyle\sim}}
$}}}}
\title{Wide-field 1-2~GHz research on galaxy evolution -- synergies with multi-wavelength surveys}

\ShortTitle{Synergies with multi-wavelength surveys}

\author{\speaker{Matt J.~Jarvis}\thanks{RCUK Fellow}\\
        Centre for Astrophysics Research, STRI, University of Hertfordshire, Hatfield, AL10 9AB\\
        E-mail: \email{M.J.Jarvis@herts.ac.uk}}

\abstract{In these proceedings I discuss various extragalactic surveys
  which will be undertaken over the next few years and which will be
  complementary to any H{\sc i} and/or continuum surveys with the
  SKA-precursor telescopes. I concentrate on the near-infrared public
  surveys which will be undertaken with the Visible and Infrared
  Survey Telscope for Astronomy (VISTA), and in particular the VISTA
  Deep Extragalactic Observations (VIDEO) survey which will provide
  the ideal data set to combine with any deep SKA-precursor
  observations of the extragalactic sky. After highlighting the links
  that the SKA precursors have with the various VISTA surveys, I
  briefly describe two forthcoming {\em Herschel} surveys, {\em Herschel}-ATLAS survey and HerMES which have a large
  scientific overlap with the SKA-precursor telescopes. Finally, I
  present a case study in combining
  multi-wavelength data sets with radio-frequency surveys to find the
  highest redshift radio sources with the aim of probing the epoch of reionization.}

\FullConference{Panoramic Radio Astronomy: Wide-field 1-2 GHz research on galaxy evolution - PRA2009\\
		 June 02 - 05 2009\\
		 Groningen, the Netherlands}

\begin{document}

\section{Introduction}

The Square Kilometre Array (SKA) promises to be one of the most
influential telescopes of any era in astronomy. In the area of galaxy
formation and evolution, it will be able to map out the neutral
Hydrogen content of galaxies up to and possibly beyond $z \sim
1.5$. However, it will be insensitive to the underlying stellar
populations, the nebula emission and the amount of dust along with various physical mechanisms that are associated with galaxies which do not emit at radio wavelengths. Thus, even in the age of the SKA, complementary multi-wavelength surveys will still be paramount to further our understanding of the Universe. Furthermore, the various SKA precursor telescopes will also benefit greatly from multi-wavelength surveys which are being or will be carried out concurrently with the deep radio surveys. 

Such surveys are already underway or being planned, and many more will
undoubtedly be initiated in the coming years. In these proceedings I
focus on the near-infrared surveys which will be carried out by the
ESO-Visible and Infrared Survey Telescope for Astronomy (VISTA), to
begin operation in 2009 at Paranal, Chile, in particular the second
deepest tier of the extragalactic public surveys to be undertaken on
this telescope, namely the VISTA Deep Extragalactic Observations
(VIDEO) survey. I also discuss the various other multi-wavelength
surveys that will be carried out over these regions of sky.

I also discuss two surveys to be conducted
with the {\sl Herschel Space Observatory}, {\em Herschel}-ATLAS and HerMES, which I believe will be of
benefit and benefit from observations with the various SKA-precursor
telescopes.

\section{Extragalactic near-infrared surveys with VISTA}\label{sec:vistasurveys}

VISTA will conduct several public surveys at near-infrared wavelengths ($0.8 - 2.4\mu$m). There are currently six approved surveys, I briefly discuss three of the surveys most relevant for extragalactic studies with the SKA pathfinders and in section~\ref{sec:VIDEO} I discuss in more detail the VIDEO survey\footnote{Further details of all of the public surveys to be conducted on both the VISTA and VST can be found in Arnaboldi et al. 2007.}: 

\noindent
$\bullet$ Ultra-VISTA is the deepest tier of the extragalactic surveys. It will survey the COSMOS field with three separate strategies. The ultra-deep survey covers 0.73~deg$^{2}$ to AB magnitudes of $Y=26.7$, $J=26.6$, $H=26.1$ and $K_{s}=25.6$, with the aim of detecting galaxies within the epoch of reionisation at $z>6$. The narrow-band survey is expected to find $\sim 30$ Ly$\alpha$ emitters at $z \sim 8.8$ to a depth of NB$_{\rm AB}=24.1$ and the wide survey will complete the coverage of the full 1.5~deg$^{2}$ COSMOS field, to depths of  $Y=25.7$, $J=25.5$, $H=25.1$ and $K_{s}=24.5$ (all AB).

\smallskip

\noindent
$\bullet$ The VISTA Kilo-degree Infrared Galaxy (VIKING) survey aims to survey two stripes at high galactic latitude in five near-infared filters. The areas have been chosen to overlap with the regions of sky covered by the 2df galaxy redshift survey (Colless et al. 2001) and the optical Kilo-Degree Survey (KIDS) to be conducted with the VLT Survey Telescope, thus providing both spectroscopic redshifts up to $z\sim 0.3$ and photometric redshifts up to and above $z\sim 1$ with typical uncertainties of $\Delta z/(1+z) \sim 0.1$ from this 9-band survey. The science aims are heavily based on this photometric redshift accuracy for cosmological studies to constrain the dark energy component in the Universe, gain more accurate mass density measurement from weak lensing, find $z>7$ quasars and trace galaxy evolution from $z\sim 1$ to the present day. These regions are also the subject of the {\em Herschel} large-area survey (see section~\ref{sec:H1K}).
It is apparent that any $\sim 1000$~deg$^{2}$ survey from the SKA-precursors should target these regions to maximise the scientific productivity of both continuum and H{\sc i} surveys.
\smallskip

\noindent
$\bullet$ The VISTA Hemisphere Survey (VHS) aims to survey the {\it rest} of the southern sky which is not being covered by the other VISTA surveys. Although much shallower than the other surveys it will provide data covering $\sim 18000$~deg$^{2}$ to a depth of at least $K_{\rm s} \sim 19.8$ and $J\sim 20.9$.
The majority of this area will alslo be covered by various surveys conducted at optical wavelengths, namely the Dark Energy Survey (DES; https://www.darkenergysurvey.org/) and the VST-ATLAS survey, in addition to the various galactic plane studies. Such a survey has wide-ranging scientific goals and its complementarity to any large area SKA precursor survey is obvious. If the SKA precursors progress as planned toward the full SKA then it is plausible that H{\sc i} redshifts could be obtained for a significant fraction of the VHS galaxies, allowing detailed investigations between the H{\sc i} and the stellar properties of the galaxies.
\smallskip

\begin{center}
\includegraphics[width=13.5cm]{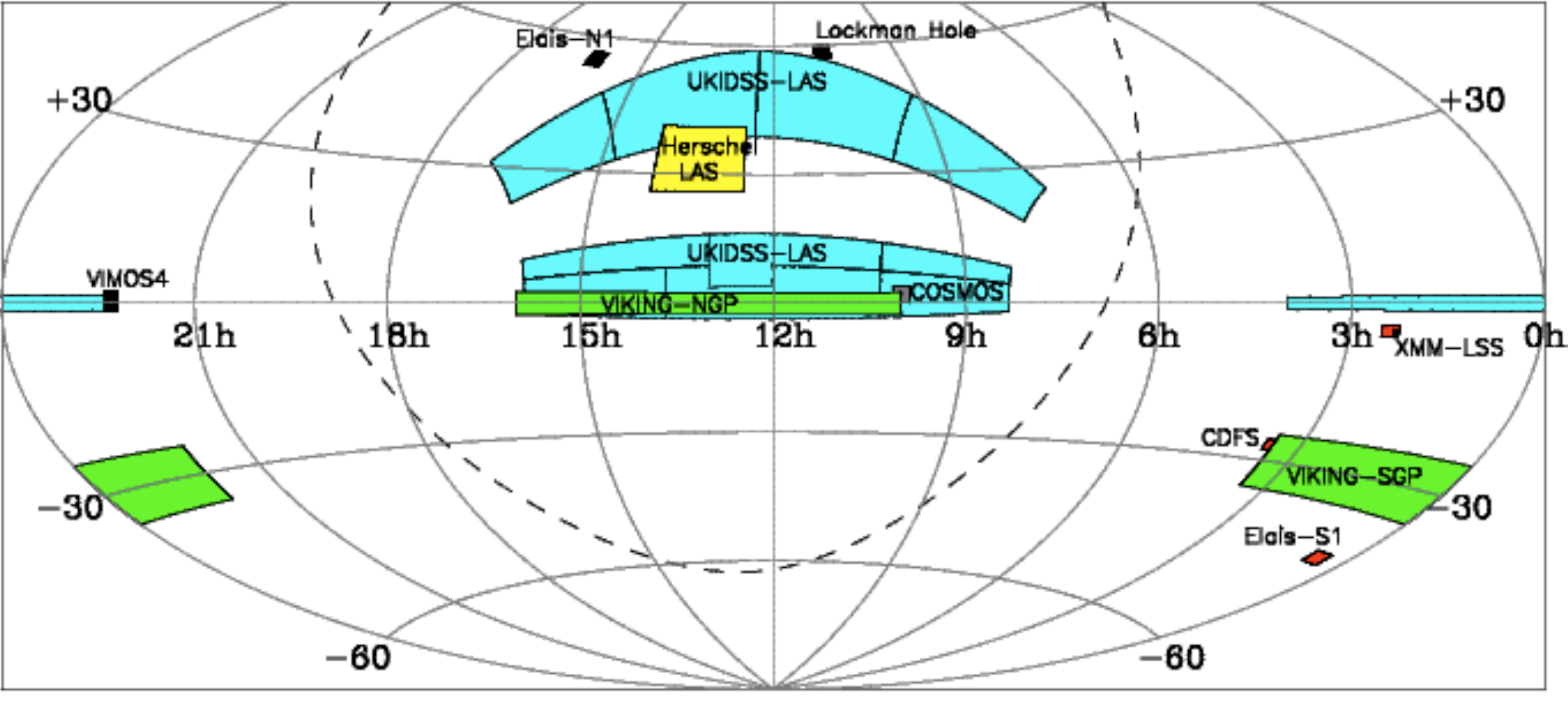}

{\textsf {\bf Figure~1.} Overview of the various VISTA surveys described in the text along with the regions to be observed with the UK Infrared Deep Sky Survey (UKIDSS; www.ukidss.org) and the Herschel large-area survey (Herschel-LAS). (Best viewed in colour.)}
\end{center}

\section{VISTA Deep Extragalactic Observations (VIDEO) survey}\label{sec:VIDEO}

In this section I describe the VISTA Deep Extragalactic Observations
(VIDEO) survey. VIDEO aims to obtain both deep and wide near-infrared
observations over well-studied contiguous fields covering
$3-4.5$~deg$^{2}$ each. The fields chosen for this are derived from
the equatorial and southern fields of the Spitzer Wide-area Infrared
Extragalactic Survey (SWIRE; Lonsdale et al. 2003), namely $\sim
3$~deg$^{2}$ within the Elais-S1 region, $\sim 4.5$~deg$^{2}$ within
the XMM-Newton Large Scale Structure (XMM-LSS) survey and 4.5~deg$^{2}$
around the Chanda Deep Field South (CDFS).  A recently approved companion survey, the {\em Spitzer} Extragalactic Representative Volume survey (SERVs), has also been allocated 1400~hours with the {\em Spitzer-}warm mission. These two surveys will enable galaxy evolution to be traced from the epoch of reionization through to the present day and as a function of environmental density. 

The VIDEO survey is planned to reach the following $5\sigma$ (2~arcsec
point source) depths; $Z=25.7$, $Y=24.6$, $J=24.5$, $H=24.0$ and
$K_{\rm s}=23.5$ (all AB magnitudes), and will use just over
200~nights of observing time over the next five years. In terms of the
galaxies which will be detected in the VIDEO survey, these depths
correspond to an $L^{\star}$ elliptical galaxy up to $z\sim4$ and
$0.1~L^{\star}$ galaxy up to $z\sim1$, thus it sits naturally between
the Ultra-VISTA and VIKING surveys.The SERVS depths are slightly
deeper in real terms with the proposed depth being able to detect a
$L^{\star}$ elliptical galaxy up to $z\sim 5$.

This combination of depth and area ensures that with the VIDEO survey
and SERVS we will be able to trace galaxy formation and evolution over the majority of cosmic history and over all environmental densities. Some of the main science goals are; 

\noindent
$\bullet$ to trace the evolution of galaxies from the earliest epochs until the present day. The depth and breadth of the VIDEO survey will enable galaxy properties to be measured as a function of environmental density,  allowing the detailed study of how the the environmental richness may affect the properties of the galaxies. Furthermore, the combination of VIDEO data with the wealth of multi-wavelength data over these regions of sky, will allow the star-formation rate and AGN activity to be be measured and linked to the stellar mass in the galaxies.
\smallskip

\noindent
$\bullet$ to measure the clustering of the most massive galaxies at $z>5$ where, extrapolating from the results of McLure et al. (2006), we expect to detect around 300 galaxies with $M>10^{11}$~M$_{\odot}$ and around 150 galaxies at $z>6$.
\smallskip

\noindent
$\bullet$ to trace the evolution of galaxy clusters from the formation epoch until the present day (see e.g. van Breuekelen et al. 2006, 2007). The depth of VIDEO ensures that we will be able to trace the bright end of the cluster luminosity function to $z\sim 3$, while its area should provide a sample of $\sim 70$ clusters with $M>10^{14}$M$_{\odot}$ at $z>1$, with around 15 of these expected to be at $z>1.5$.
\smallskip

\noindent
$\bullet$ to quantify the accretion activity over the history of the Universe. The depth and area, along with the filter combination will allow the detection of the highest redshift quasars, and this place the first constraints on the quasar luminosity function at $z>6.5$, when combined with  wider and shallower surveys, such as VIKING. Furthermore, VIDEO data can be combined with Spitzer and {\em Herschel} data to place constraints on both the obscured AGN and star-forming galaxies, all the way out to $z\sim6$ if AGN host galaxies have typical luminosities of $2-3~L^{\star}$ (e.g. Jarvis et al. 2001a).


The VIDEO survey will therefore provide the ideal data to be combined with any deep extragalactic observations with the SKA-pathfinder telescopes in both continuum and H{\sc i}. Crucially, VIDEO will have the depth to detect low-mass galaxies which would dominate the $\ltsim1$ SKA H{\sc i} surveys, thus providing important information on the stellar populations in these galaxies.

\section{Herschel \& SCUBA-2}

To obtain a full picture of the evolutionary history of the Universe one needs to combine data from all wavelengths which are emitted by a range of processes. Therefore, the forthcoming power of {\em Herschel} and SCUBA-2 will provide the crucial measurement of the obscured activity in the Universe. 

\subsection{{\em Herschel}-ATLAS}\label{sec:H1K}
{\em Herschel}-ATLAS (H-ATLAS)  obtained the largest time allocation ($\sim 600$~hours) of any programme in a call for Open Time Key Projects and will begin observations in Autumn 2009, it will cover $\sim 550$~square degrees in five far-infrared and submm wavbands from 100-550$\mu$m.  H-ATLAS has scientific aims that span star-forming regions in our own galaxy through to the most distant accreting black holes. Briefly summarising the details of this survey and highlight some of the scientific goals;

\noindent
$\bullet$ The H-ATLAS will be the first submm survey large enough to detect a
significant number of galaxies in the nearby universe, between 22,000
and 77,000 individual galaxies out to $z \sim 0.3$.  By carrying out the
survey in the fields surveyed in the SDSS, 2dFGRS and the new GAMA
survey (Driver et al. 2009), we estimate that $\approx 50$\% will
already have redshifts, including $\approx$95\% of those at $z < 0.1$.

\smallskip
\noindent
$\bullet$ H-ATLAS will be used to investigate the relationship
between the star formation and the black hole activity and how this
relationship changes over time by observing a very large sample of
quasars drawn from the SDSS.

\smallskip

\noindent
$\bullet$ The strong negative $k-$correction at submm wavelengths means that large area shallow surveys such as H-ATLAS provide an excellent way of finding large numbers of strong gravitational lenses. Models predict that the H-ATLAS will contain $\sim 1500, 800$ and $350$ strongly-lensed galaxies at $250, 350$ and $500\mu$m, respectively, with a lens yield ranging from 1\% at 250$\mu$m to close to 100\% at 500$\mu$m 
\smallskip

\noindent
$\bullet$ H-ATLAS will detect $\sim220,000$ sources with a median redshift of $\sim 1$ and will therefore contain a large amount of information about large-scale structure upto a scale of 1000~Mpc at $z\sim 1$. With this it will be possible to measure the angular correlation function of 
the sources on very larges scales, allowing estimates of the masses of the 
dark-matter halos and discriminate between competing models of galaxy formation.
\smallskip

\noindent
$\bullet$ H-ATLAS will make the most sensitive survey yet carried out of Galactic dust at high latitudes, allowing the identification of prestellar and protostellar cores over a wide area down to the Jupiter mass regime and to determine the fraction of isolated high-latitude star formation versus the more well-known clustered mode.


The equatorial ATLAS fields are those covered by the Galaxy and Mass Assembly (GAMA) survey which will provide $\sim 250,000$ redshifts over these fields from AAOmega along with the VISTA-VIKING and VST-KIDS data sets, we will have a unique data set with which to study galaxy and AGN evolution in the $z<1$ Universe.

\subsection{HerMES \& SCUBA-2 Cosmology Legacy Survey}
It is also important to emphasize that the $z>1$ data set over the VIDEO/SERVs fields will have a similar range in multi-wavelength data to that of the {\em H-}ATLAS/VIKING/KIDS fields, i.e. they will be covered by {\em Herschel} and SCUBA-2.
The {\em Herschel}-Guaranteed Time survey {\em Her}mes will survey two of the fields covered by VIDEO to much deeper levels than {\em H}-ATLAS, allowing a view of the obscured activity at $z>1$. This will allow the detailed study of both the obscured and unobscured build-up of stellar and black-hole mass over the history of the Universe.
The SCUBA-2 Cosmology Legacy Survey (S2CLS) will also survey the VIDEO/SERVs fields. It is worth noting that the VIDEO and SERVs depths are such that they will be able to detect many of the SCUBA sources and future radio facilities such as LOFAR, ASKAP and MeerKAT should detect all of them.

Therefore, the combination of the multi-wavelength data from the $\sim 100$~square degree fields such as H-ATLAS, VIKING and KIDS with the deeper, smaller fields such as HerMES, VIDEO and SERVS will provide a powerful data set for understanding the evolution of galaxies across cosmic time and as a function of environmental density.

\section{A Case Study in finding high-redshift radio sources with
  multi-wavelength surveys}

As an example of what can be achieved by combining multi-wavelength data sets we describe a new method for finding the highest-redshift radio galaxies (HzRGs). 
HzRGs (by which we mean those at $z>4$) represent only a very small fraction of all radio sources at a given
flux-density limit and it is necessary to filter out low-redshift contaminants
before performing spectroscopy. This has previously been done by applying a filter based on radio
properties such as steep spectral index (e.g. Chambers et al. 1996; De Breuck et al, 2002; Cohen et al. 20046). The next stage
is to take $K$-band images of the targets, since radio galaxies follow
an extremely tight locus in the $K$-band Hubble diagram (Jarvis et al. 2001a) and therefore HzRGs will be faint. 
However,
although $K$-band imaging is very reliable at identifying radio
galaxies, these observations are very expensive, needing to reach
$K\sim20$ to detect sources at the epoch where the radio-AGN density peaks
($z\sim2$; Jarvis \& Rawlings 2000; Jarvis et al. 2001b). Unfortunately, even after filtering on radio source properties, such as angular size and spectral index, the fraction of
HzRGs remains low -- e.g., 1/68 sources in the 6C** sample of Cruz et
al.\ (2006, 2007) lies at $z>4$ -- so most of the follow-up near-infrared imaging 
simply rules out high-redshift targets. 

The recent advent of wide-field deep
near-infrared surveys has opened up a new path for finding such objects. The
{\em Spitzer}-SWIRE (Lonsdale et al. 2003) and the UKIDSS Deep Extragalactic Survey (DXS; see e.g. Warren et al. 2007) are deep enough to reliably
eliminate all $z<2$ radio galaxies. Therefore by
targeting only radio sources which are very faint or not detected in the {\em Spitzer}-IRAC channels 1 and 2 or the
DXS $K$-band imaging, we are likely to be probing radio sources in the 3\,Gyr after the Big Bang.

We have cross-matched the {\em Spitzer}-SWIRE data with sources at $>10$mJy from the 1.4~GHz FIRST survey in the Lockman Hole, Elais-N1 and Elais-N2 fields, which together cover a total of $\sim 24$~square degrees.  For our initial search we use a 3.6$\mu$m limit of $<30\mu$Jy in a 1.9~arcsec radius aperture, as given in the SWIRE catalogues. 
Our choice of radio flux-density limit ensures that all objects at $z>2$
are above the break in the radio luminosity function and will
therefore possess strong emission lines, particularly Ly$\alpha$ which is redshifted into the optical visible part of the spectrum, allowing redshift identification in much shorter integration times than if signal-to-noise on the continuum were needed. 
Simulations (Jarvis \& Rawlings 2004; Wilman et al. 2008)
predict that $\sim$1 in 15 of these sources will lie at $z>4$.
This flux-density limit also ensures that the number of lower-redshift lower luminosity radio sources are also reduced significantly with only $\sim 1$ in 10  expected above our flux-density limit in 24 square degrees. The survey will be described in more detail in a forthcoming paper (Teimourian et al. in prep.).

Our first results from this study are presented in Jarvis et al. (2009) but we summarize our findings here. The radio source J163912.11+405236.5 is detected in the FIRST survey with a flux-density of 22.5~mJy and is unresolved at the 5~arcsec resolution of this survey. 
There is also a source detected in the 325~MHz Westerbork Northern Sky Survey (WENSS; Rengelink et al. 1997) at 16 39 12.17 +40 52 40.3 (J2000) which is 3.6~arcsec away from the FIRST centroid, therefore we associate this source with the FIRST source. The WENSS catalogue gives a flux-density of $67\pm5$~mJy for this source. Assuming a power-law spectral index between 325~MHz and 1.4GHz the spectral index of the source is therefore $\alpha = 0.75\pm0.05$\footnote{We use the convention for spectral index $S_{\nu} \propto \nu^{-\alpha}$}. Thus this source would not fall into the category of ultra-steep spectrum sources which have been used to search for high-redshift radio galaxies in recent years. In figure~2 we show the 3.6$\mu$m image from the {\em Spitzer}-SWIRE survey overlaid with the radio image from the FIRST survey. There is a very faint source in the 3.6$\mu$m image at the centre of the radio position which we identify as the host galaxy but with no detection in any of the other SWIRE-IRAC or MIPS-24$\mu$m data. 30~minutes of optical spectroscopy with the WHT-ISIS spectrograph revealed a bright Lyman-$\alpha$ emission line at a redshift of $z=4.88$ for this source (Fig.~3), the second highest redshift radio galaxy found to date. This is particularly impressive given that this only used 5~hours of observing time with the WHT to find this object and many the other HzRGs at $z>3$ (Teimourian et al. in prep.).

This therefore shows the efficiency of combining deep multi-wavelength data which already exists rather than adopting the traditional follow-up strategy for radio surveys, where the fields have been distinct from where the wealth of multi-wavelength data has been.

\begin{minipage}[h]{\linewidth}
\begin{minipage}[h]{0.45\linewidth}
\includegraphics[width=0.9\columnwidth]{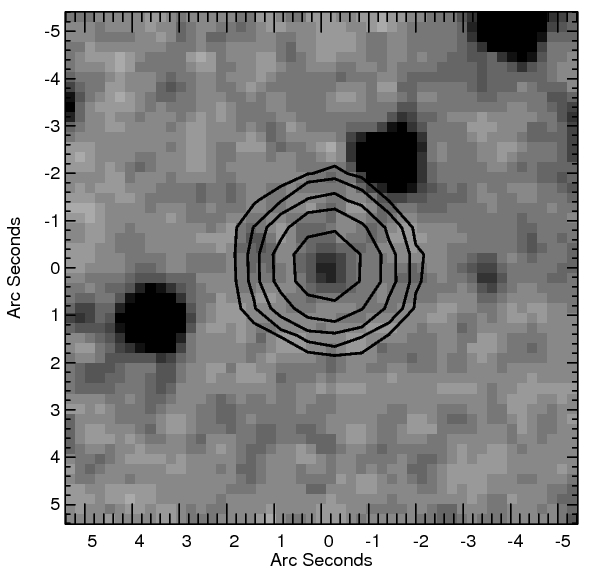}

{\textsf {\bf Figure~2.} The {\em Spitzer}-3.6$\mu$m image (greyscale) overlaid with radio contours from the FIRST survey. Contour levels are 0.8, 1.6, 3.2, 6.4 and 12.8~mJy/beam.}
\label{fig:overlay}
\end{minipage}
\hspace{0.5cm}
\begin{minipage}[h]{0.45\linewidth}
\includegraphics[width=\columnwidth]{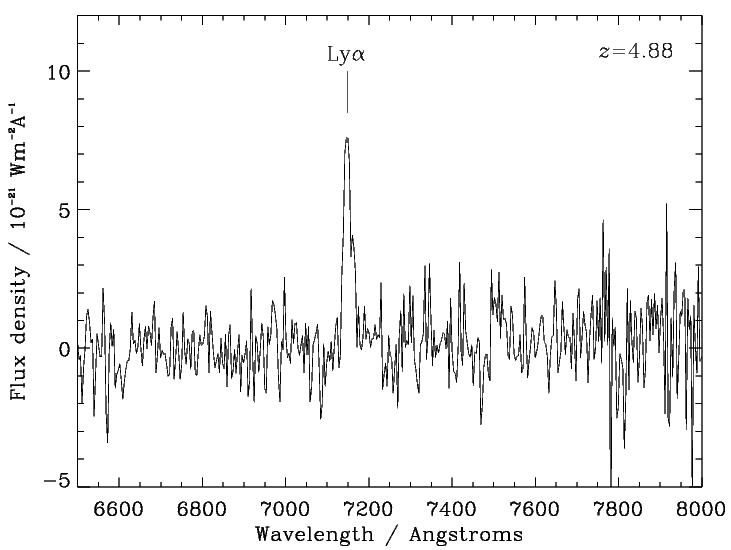}

{\textsf {\bf Figure~3.} 1-dimensional spectrum of the radio galaxy  J163912.11+405236.5. One can see the bright Lyman-$\alpha$ emission line at $\lambda=7149$\AA, which corresponds to a redshift of $z=4.88$. }
\label{fig:spectrum}
\end{minipage}
\end{minipage}

\section{Summary}

Over the next few years we will be in the unique position of having large multi-wavelength data sets over the same patches of sky, from medium-depth surveys covering $\gtsim 100$~square degrees to very deep surveys covering $1-10$~square degree areas. The wealth of science that could be gleaned from combining these data sets is huge and the future radio surveys would benefit enormously by also surveying these fields. With radio continuum and H{\sc i} observations from the SKA-precursor telescopes, we could soon be in a position to understand all of the crucial components of galaxies.

\section{Acknowledgements}
I would like to thank the various PIs of the surveys which I discuss for actually leading the surveys. Thanks also to Simon Dye who provided the code to generate figure~1 and my collaborators on the search for high-redshift radio galaxies, Hanifa Teimourian, Chris Simpson, Dan Smith, Steve Rawlings and Dave Bonfield.

\end{document}